\documentclass[final,3p,times,twocolumn]{elsarticle}

\usepackage{graphicx}

\usepackage{amssymb}

\usepackage{lineno}

\biboptions{numbers,square,sort&compress}

\usepackage{color}

\def\aver#1{<\!\!#1\!\!>}

\journal{RICH2010 Proceedings, Nuclear Instruments and Methods A}

\begin{document}

\begin{frontmatter}



\title{Cherenkov Light Imaging -- Fundamentals and recent Developments}


\author{J\"urgen Engelfried}
\ead{jurgen@ifisica.uaslp.mx}
\address{Instituto de F\'{\i}sica, 
Universidad Aut\'onoma de San Luis Potos\'{\i}, Mexico}

\begin{abstract}
We review in a historical way the fundamentals of Cherenkov light imaging
applied to Ring Imaging Cherenkov Counters.
We also point out some of the newer developments in this very active field.
\end{abstract}


\end{frontmatter}



\section{Introduction}

In the early 1930's Pavel A.~Cherenkov, under the guidance of his adviser
Sergey I.~Vavilov, investigated the nature of light emitted by radioactive
minerals.  Eventually it became clear that he observed a new phenomena,
later called {\sl Cherenkov} or {\sl Vavilov-Cherenkov} radiation.
The electrons from the $\beta$-decay with a velocity~$v$ larger than the 
velocity of light in a medium (in Cherenkov's case water) with
refractive index~$n$
emit light under
a fixed (Cherenkov)-angle $\theta_C$ given by~\cite{Cherenkov}
\begin{equation}
\cos\theta_C = {{1}\over{\beta\, n}} = {{1}\over{{{v}\over{c}}\, n}}
\label{changle}
\end{equation}
Shortly afterwards Ilja M.~Frank and Igor Y.~Tamm brought forward 
a theory explaining not only the emission angle, but also the intensity
and wavelength (or energy) dependence~\cite{francktamm}:
\begin{equation}
   {{d^2N}\over{dE\,dl}} = {{\alpha z^2}\over{\hbar c}}
\left(1 - {{1}\over{(\beta n)^2}}\right)
= {{\alpha z^2}\over{\hbar c}} \sin^2\theta_C
\label{franktamm}
\end{equation}
Cherenkov, Frank, and Tamm shared the 1958 Nobel Price for this discovery.

\section{Fundamentals of Ring Imaging}

The basic components of a detector or detector system utilizing the 
Cherenkov effect are: (1)~a radiator (2)~a photon detection system,
and (3)~optional some focusing elements like mirrors or lenses.
Examples of Cherenkov detectors include water tanks lined with 
large photomultipliers like SuperKamiokande~\cite{Fukuda:2002uc} 
or SNO~\cite{Boger:1999bb},
working in an effective Proximity Focusing mode; Ring Imaging Cherenkov
Counters (RICH) with gas radiator and mirrors 
like in SELEX~\cite{Engelfried:1998tv}, 
and DIRC (detection of internally reflected
Cherenkov light) counters like in BaBar~\cite{Adam:1999zt}.

The first mentioning (altough not with this name) of a Ring Imaging
Cherenkov Counter is an article by Arthur Roberts 
``A new type of Cherenkov detector for the accurate measurement
of particle velocity and direction'', 
published in 1960~\cite{Roberts:1960zz}. 
The abstract reads:
\begin{quotation}
{\sl \ldots The image is a ring, whose diameter measures accurately the
Cherenkov cone angle, and thus the particle velocity. In addition the
coordinates of the center of the circular image accurately indicate the
orientation of the particle trajectory (though not its position)\ldots}
\end{quotation}

Only in the 1970's larger area photon detection became possible thanks
to the work of 
Tom~Ypsilantis and Jaques~Sequinot~\cite{Seguinot:1976yp}, using TEA
and later TMAE as photo-sensitive agents within multiwire- or drift-chambers
to detect the released electron(s).

In addition to the huge PMT's employed in the Water Cherenkov detectors,
during the 1980's small photomultipliers (as small as $13\,\mbox{mm}$ 
diameter)
were employed in RICH detector
prototypes in Protvino~\cite{Ronzhin:1986dq}.

\subsection{Ring Imaging Cherenkov -- The Basics}

\begin{figure}[htb]
\centerline{\includegraphics[width=0.49\textwidth]{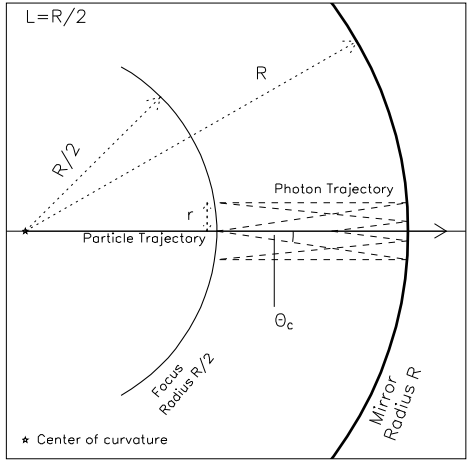}}
\caption{Working prinziple of a RICH detector. Photon trajectories
are drawn in the small angle approximation.}
\label{fig:richprinzip}
\end{figure}
\begin{figure}[htb]
\centerline{\includegraphics[bb=10 165 555 400, clip, width=0.49\textwidth]{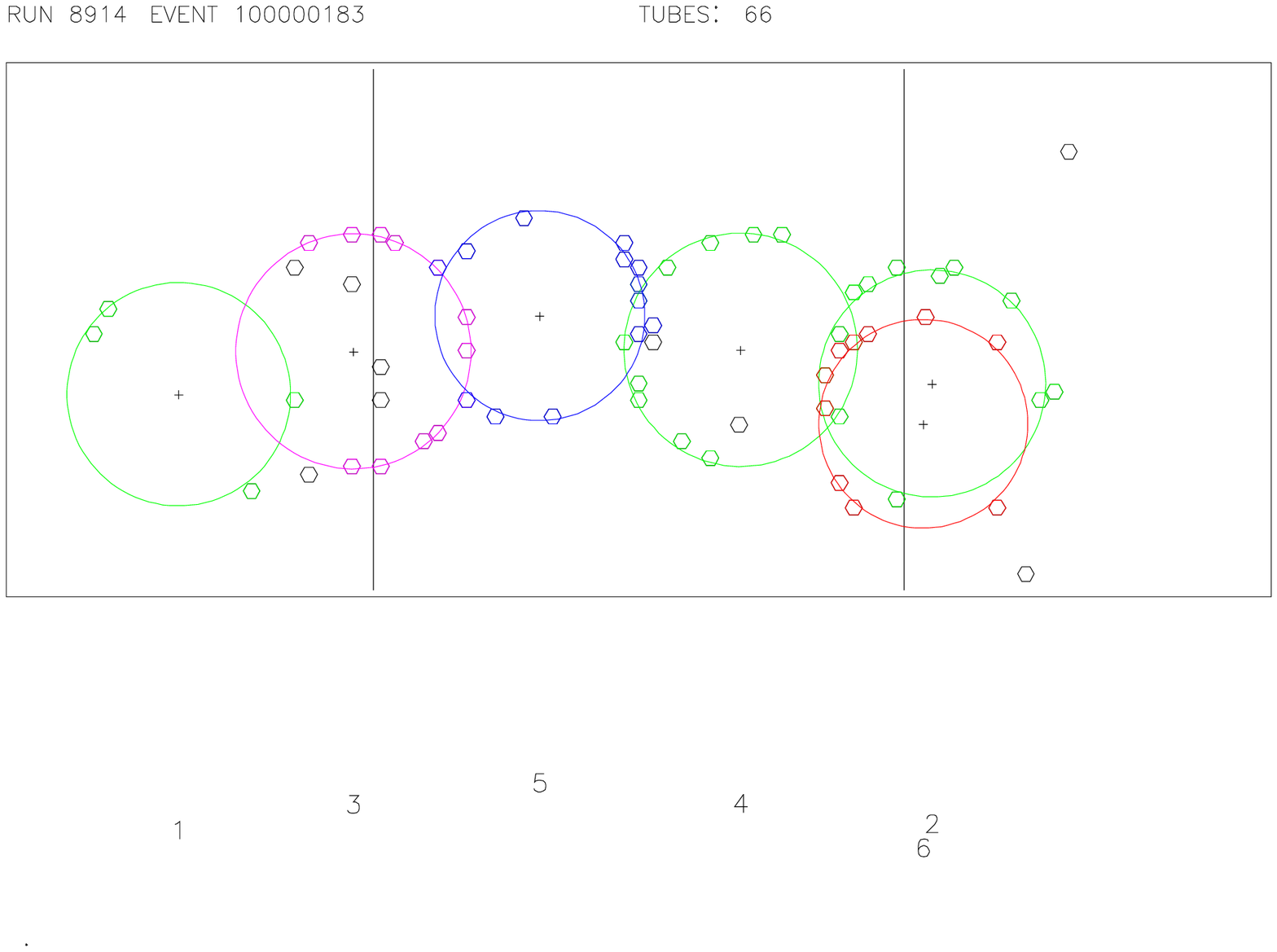}}
\caption{Typical event in the SELEX RICH.
The small hexagons represent hit phototubes.The rings are drawn with
the radius of the most likeli mass hypothesis from a
likelihood analysis~\cite{Muller:1993ig}, and color coded. 
From left to right: $\pi^+$, $e^+$, $p$, $\pi^-$. $K^-$, $\pi^-$.
The detector has 2848 half inch photomultipliers, covering a total area
of $146\,\mbox{cm}\times54\,\mbox{cm}$.
Please note the low noise -- only two hits are not assigned -- and compare
to fig.~\ref{wa89onl}.}
\label{selexevent}
\end{figure}

As shown in fig.~\ref{fig:richprinzip} the Cherenkov photons emitted under
an angle $\Theta_C$ with respect to the particle trajectory in a radiator with
refractive index~$n$ are focused (in the small angle approximation) 
by a spherical mirror of radius $R$ onto a focal sphere of radius~$R/2$
(thus the focal length $F=R/2$) where
the photons form a ring of radius~$r$ given by
\begin{equation}
r = F \cdot \tan\Theta_C = {{R}\over{2}} \cdot \tan\Theta_C
\label{ringrad}
\end{equation}
while $\Theta_C$ is given by equation~\ref{changle}.
The number of detected photons~$N_{\rm ph}$ is usually expressed in the form
\begin{equation}
N_{\rm ph} = N_0 \cdot L \cdot \sin^2\Theta_C
\end{equation}
where $L$ is the radiator length (usually $L=F$, but folding the light
path with flat mirrors is possible), and $N_0$ is the ``figure of merrit''
describing the quality and wavelength interval of the photon detection 
system, first introduced in~\cite{Seguinot:1976yp}.
In fig.~\ref{selexevent}.
the final ring images in a multitrack event are shown\footnote{This image
was used within the RICH2004 conference poster.}.

In the small angle approximation, equation~\ref{ringrad} yields
\begin{equation}
r = F\sqrt{2 - {{2}\over{n}}\sqrt{1+{{m^2 c^2}\over{p^2}}}}
\label{radformula}
\end{equation}
where $m$ is the mass and $p$ is he momentum of the particle producing
the Cherenkov light.
Thus, knowing $p$ (measured in a magnetic spectrometer) $m$ can be obtained,
e.g.\ the particle can be identified; also possible is the opposite, knowing
(or assuming) $m$ $\beta$ or $p$ can be obained.

\begin{figure}[htb]
\centerline{\includegraphics[width=0.49\textwidth]{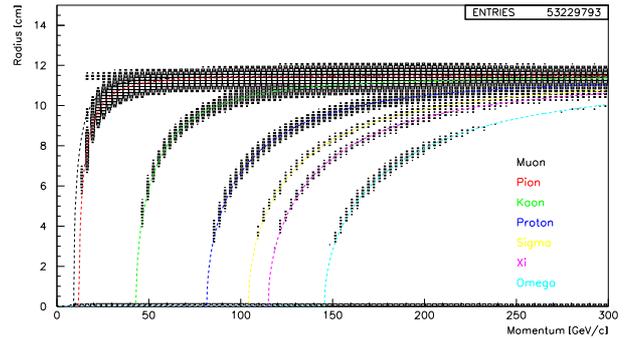}}
\caption{Radius versus momentum for more than 53~million positive tracks
in the SELEX RICH~\cite{Engelfried:2002mh}. The dashed lines are absolute
predictions for the different mass hypotheses according to 
equation~\ref{radformula}.}
\label{selexradiusdist}
\end{figure}

The first RICH detector showing a scatter plot of ring radius versus momentum
with real experiment data is the E605 RICH detector~\cite{Adams:1984ac}
in 1983, with $\aver{N_{\rm ph}}=2.7$, where the curves for $\pi$ and $K$ 
can barely be seen.
20~years later, in 2003, SELEX published~\cite{Engelfried:2002mh} a plot 
with all 16~stable particles clearly visible in the scatter plot.
Figure~\ref{selexradiusdist} shows the plot for postive tracks.

The angular separation between two particles 
of masses $m_1$ and $m_2$ is given by
\begin{equation}
\Theta_C \Delta\Theta_C = {{m_1^2-m_2^2}\over{2p^2}}$$
\label{eqn:separ}
\end{equation}
In hadro-production experiments this means that usually pions and kaons
are the particles which are most difficult to separate in a RICH detector,
because electrons and muons are identified by other means; but really 
the smallest separation is between muons and pions, which is actually
used as a key feature in the velocity spectrometer concept pioneered by
CKM~\cite{Engelfried:2002xh,Cooper:2004hz} and used in
NA62~\cite{Bucci:2008zza,Lenti:2010}.

After all these preliminary statements it looks like that a 
RICH Detector is as simple as a radiator box, some mirrors, 
and a few phototubes.  This is obviously not the case, and we will elaborate
on this in the following parts.

\subsection{Short History of RICHes in Experiments}
As mentioned earlier, a first generation of 
RICH detectors were employed in experiments
in the early 1980's. Examples include Fermilab E605~\cite{Adams:1984ac},
experiments WA69~\cite{Buys:1996gd} and WA82~\cite{Siebert:1993di}
at the CERN Omega-Spectrometer, and Fermilab
E665~\cite{Coutrakon:1988nu}. These
detectors suffered from operational problems,
and only had a limited contribution
to the physics results of the experiments.

Second generation detectors, based on the (positive and negative) experiences
gained earlier, were used at the end of the 1980's and beginning of 90's.
Examples include an upgraded Omega RICH~\cite{Siebert:1993di} used by
WA89~\cite{Muller:1999rj} and WA94~\cite{Abatzis:1996ga},
Delphi~\cite{Adam:1994yp},
SLD-GRID~\cite{Abe:1993ku,Va'vra:1999rh}, 
CERES~\cite{Baur:1993mw,Baur:1993mz}, and SPHINX~\cite{Dorofeev:1994um}.
These detectors reached about half of the originally expected
number of detected photons, but all the losses could be eventually 
accounted for by carefully determining all the operational parameter; one 
large problem was the TMAE quantum efficiency and electron
attachment length depending on the cleaning proceedure of the
TMAE~\cite{Hallewell:1993tp,Martens:1993df}. 
These detectors still presented some operational 
challanges, but generally contributed to the physics ouput
of the experiments.

The third generation, build and used during the mid-90's, 
like SELEX~\cite{Engelfried:1998tv},
Hermes~\cite{Ryckbosch:1999rn,Jackson:2003vr,Jackson:2005eu,DeLeo:2008zz}, 
and  
Hera-B~\cite{Korpar:1999rq,Korpar:2003vs,Staric:2005ev},
reached their designed performance parameters from the very beginning,
were reasonable easy and stable to operate, and made essential contributions
to the physics results.

The fourth generation detectors, like
BaBar--DIRC~\cite{Adam:1999zt,Schwiening:2005bv}, 
CLEO--III~\cite{Mountain:1999kk,Artuso:2002ya,Sia:2005vg}, 
COMPASS~\cite{Baum:1999rz,Albrecht:2005ew,Abbon:2008zzc},
are all well understood devices, do not present too much operational
problems, and are part of the physics analysis of the experiment.

Newer generation devices, like the RICHes in 
LHCb~\cite{Easo:2003vt,Easo:2005fm,Harnew:2008zz} and
ALICE~\cite{Cozza:2003vv,Gallas:2005fp,Molnar:2008pr}, are just starting
operation, and at this conference first results will be presented.
There are two more RICH projects which were cancelled due to internal
politics in the USA; we should have been able to see first results
from the BTeV~\cite{Blusk:2002vf}
and CKM~\cite{Engelfried:2002xh,Cooper:2004hz} RICHes.
There is a long list of detectors in different stages of
planning and construction, like the RICHes for BELLE~II~\cite{Iijima:2008zza}, 
NA62~\cite{Bucci:2008zza,Lenti:2010}, Panda~\cite{Fohl:2008zza}, 
CBM~\cite{Hohne:2005ec,Hohne:2008zz},
and many more.

\subsection{RICH -- The Reality}
In this section we will discuss some of the obstacles encountered while
designing and operating a RICH detector.  Most of these points are 
now (today) well understood, but earlier generation RICHes suffered with them.

\subsubsection{Detector surface}
\begin{figure}[htb]
\centerline{
\includegraphics[bb=85 506 520 763, clip, width=0.49\textwidth]
{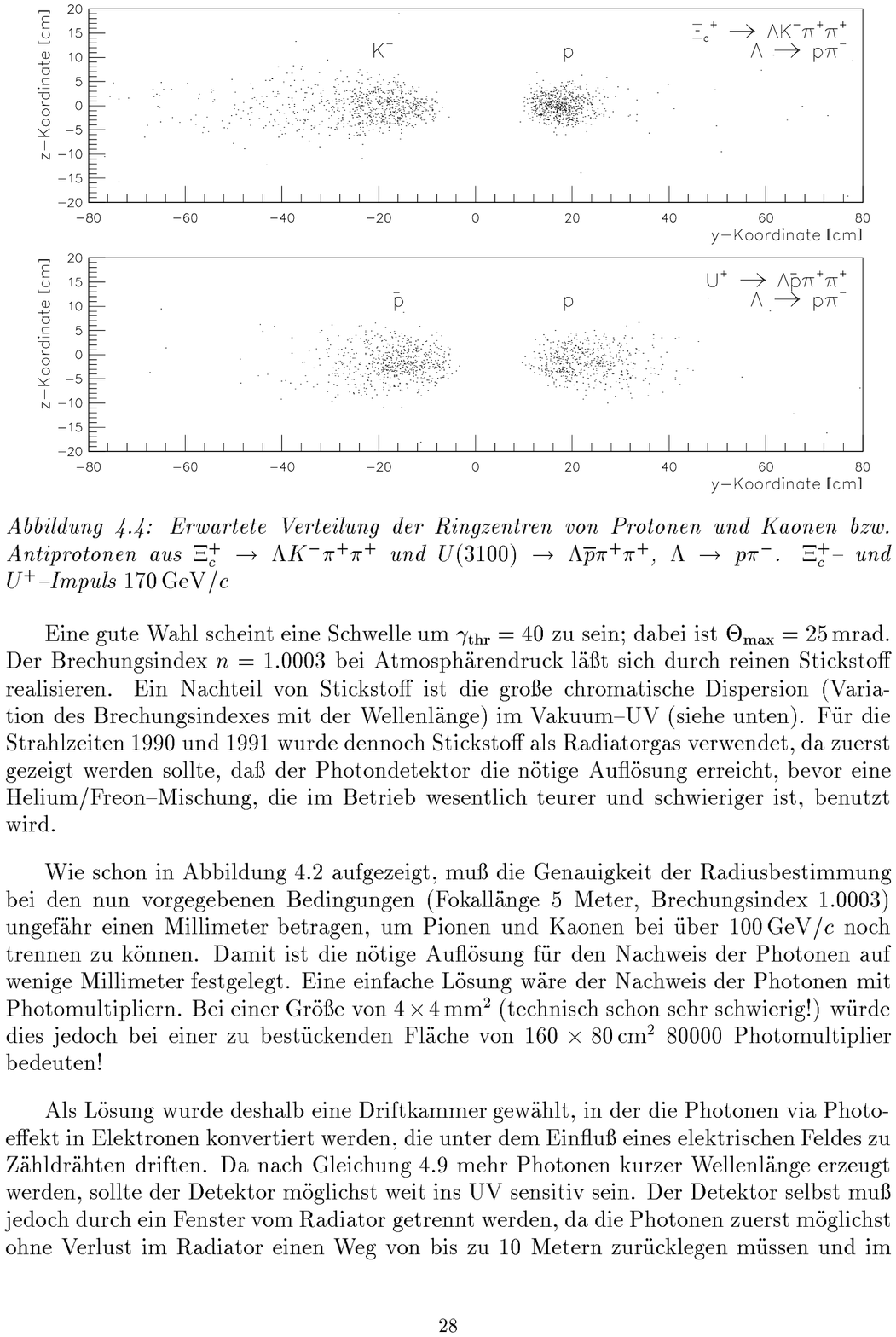}}
\caption{Distribution of expected ring centers for different final
state particles from decays:
$\Xi_c^+\to \Lambda K^-\pi^+\pi^+$,
$U^+ \to \Lambda \overline{p} \pi^+\pi^+$,
$\Lambda \to p\pi^-$, for $\Xi_c^+$ and $U^+$ momenta of $170\,\mbox{GeV}/c$. 
Simulation result for WA89 (ugraded Omega-RICH)~\cite{Engelfried:1992aaa}.}
\label{ringcenters}
\end{figure}

Obvious from fig.~\ref{fig:richprinzip} (and also mentioned by A.~Roberts)
is the fact that the ring center is given by the track angle and not
by the track position. Since most RICH detectors are placed downstream of a
magnetic field used the measure the track momentum, the photon detector
has to cover a large surface area, as shown in fig.~\ref{ringcenters}.
  But still the photon position should
be measured with sufficient accuracy (given by equation~\ref{eqn:separ}), 
so in general one ends up with a
huge number of ``pixels'' of some kind of photon detector.

\subsubsection{How to select the refracive index $n$?}

Most (but not all) photon detectors operate in the ultraviolet or
even VUV region, because due to the higher photon energy it is generally
easier to effectively detect the photon. Since the RICHes are usually
placed downstream of a magnetic field and lower momentum particles
are deflected and do not reach the RICH detector, a typical selection
for the pion threshold (and thus $n$) is the minimum momentum of the
particles reaching the detector. But also the contrary is used, for example
in the hadron-blind CERES RICH~\cite{Baur:1993mw}. 
Tables for the refractive indexes for different materials are readely available
(see for example~\cite{PDG2010}).  In case the desired $n$ is not directly
available, one can use (over- and under-)pressure or mixtures to obtain it.

One has to be extremely carefull in matching the choice of (gas-)radiator with
the sensitive wavelength range of the photon detection system. The refractive
index depends on the wavelength ($n=n(\lambda)$), and diverges close to
an absorption line of the gas, leading to a smearing of the Cherenkov angle
and to a loss of resolution (called ``chromatic dispersion'').
Early measurements of this wavelength dependence
in the VUV region were not very accurately done, but correct parametrizations
are now available and can be found in~\cite{Bideau:1982aaa,Abjean:1995up}.
At the very end one has to find an equilibrium between the number of 
photons detected (usually meaning extending the range to lower wavelengths)
and the cost of larger chromatic dispersion.

Once the refractive index (or the gas) is selected, one has to check
if the resolution at high momentum (see equation~\ref{eqn:separ}) is good
enough for the experiment. If not, there are several possible solution,
all of them adding additional complications and costs to your detecter:
smaller (and more) pixels (only if pixel size is resolution limit);
larger mirror radius (focal length), which mean larger area of the
photon detector with more pixels; more than one radiator to cover different
momentum regions (examples: Delphi, SLD, Hermes, LHCb).

In any case you should try to seal the radiator system so the refractive
index will not change over time.  This is not always possible, for example
when large windows separate the radiator from the detection volume. 

\subsubsection{Proximity Focusing}
If the radiator material has a large refractive index (like glas or liquid)
the number of photons emitted is large (equation~\ref{franktamm},
$N_{\rm ph}\propto sin^2\Theta_C$). A ``thin'' radiator yields a small
width ring; if the width is small enough for the desired resolution,
no focusing elements are needed. An extrem application of this
method are the Water Cherenkov Detectors which separate muon from
electrons without any focusing elements.

\subsubsection{Mirrors}
The mirror radius is usually twice the radiatior length, but also other
geometries with tilded flat mirrors are possible. Mirrors have to be
``thin'' (small interaction and radiation length) to avoid interactions
and $\gamma$ conversions;
but remember that the radiator itself also is material!

The surface quality has to be ``good enough'' as not to be the resolution
limit. Typically the surface quality can be worse than needed for
astronomical applications, and allows to produce thinner mirrors.
The parameters (average radius, quality) of every single mirror has to 
measured; we recommend the 
Ronchi Method~\cite{Ronchi:1923aaa,Stutte:1995us,Estrada:2005ep} to
determine a complete map of the mirror surface.

The reflectivity has to be high and has to cover the full range of the
wavelength sensitivity of the photon detection system, usually extending to
the VUV; reflectivities above $90\,\%$ at $160\,\mbox{nm}$ can be achieved.
 
Best would be a single mirror, 
but usually this is not possible due to size restrictions at the 
production or coating facilities, or the thickness and  surface quality
requirements. But several mirror segments can be mounted to form one
larger mirror, with some effort on alignment (hardware and software).
The optimum size for a mirror seems to be the radius of a $\beta=1$ 
ring; this facilitates the alignment during operation of the detector.

\subsubsection{Photon Detection System}

The general design of RICH detectors seems to be that the number of 
detected photons for a $\beta=1$ ring is around 15.  For the photon detection
we have to be aware that Cherenkov photons are single photons; noise sources
in the system have to be below the one photon equivalent.  The general
way to detect a photon is via the photoelectric effect, converting the
single photon into a single electron, or (in semi-conductor devices) into
an electron-hole pair.

Early RICH detectors used photon-sensitive vapors
like TEA or TMAE for the conversion, and a MWPC, a TPC or TEC to detect
the electron. A window separating the radiator from the photon detector
is necessary, and has to be transparent to the photons which have enough 
energy to ionize the vapor.  Also the counting gas in the chamber has to 
transparent for the relevant photons.  In the classical geometry 
(see fig.~\ref{fig:richprinzip})
the charged particle itself could pass thru the
chamber, leaving a huge $dE/dx$ signal (typically a few hundred electrons)
in the chamber and disturbing the measurement of the only photo-electron.
\begin{figure}[htb]
\centerline{\includegraphics[bb=37 137 562 439, clip, width=0.49\textwidth]{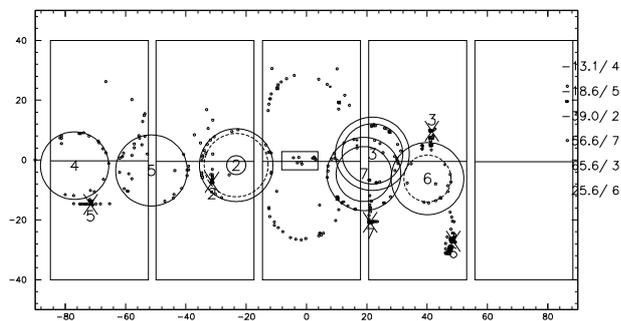}}
\caption{Single event display from 
WA89 (Upgrade Omega RICH)~\cite{Engelfried:1992aaa}. Points show hits in
the detector; circles show predicted rings for pion. kaon, and proton
hypothesis, respectively (only if above threshold); crosses show the predicted
crossing points of the tracks. The momenta of the
tracks are shown on the right.}
\label{wa89onl}
\end{figure}
An example for this is the online event display from the WA89 (upgraded Omega)
RICH in figure~\ref{wa89onl}.

The classic photon detector, the photomultiplier, is used in sizes up
to $40\,\mbox{cm}$
diameter in Water Cherenkov detectors;  and as small as 13mm in
RICH detectors. PMTs are used when the number of pixel required
is not too large; the largest PMT system (BaBar) has about 10000 channels.

Other photon detectors used or planned to be used are:
Multi-Anode PMTs,
Microchannel Plates,
Hybrids (photocathode with Silicon Strip/Pixel Detector),
CsI photocathode with a ``GEM'' to detect the electron,
and solid state (silicon) devices.
All of them have dedicated talks or full sessions within this conference.

\subsection{Contributions to total resolution}

All the parts making up the full Cherenkov detector system have some
influence on the performance, usually expressed as the single hit or
the ring radius resolution.  From there we can deduce the separation
between two particle hypotheses at a given momentum, usually expressed
in units of the ring radius resolution, which is assumed to be Gaussian.

\begin{figure}[htb]
\centerline{\includegraphics[bb= 75 508 516 760, clip, width=0.49\textwidth]
{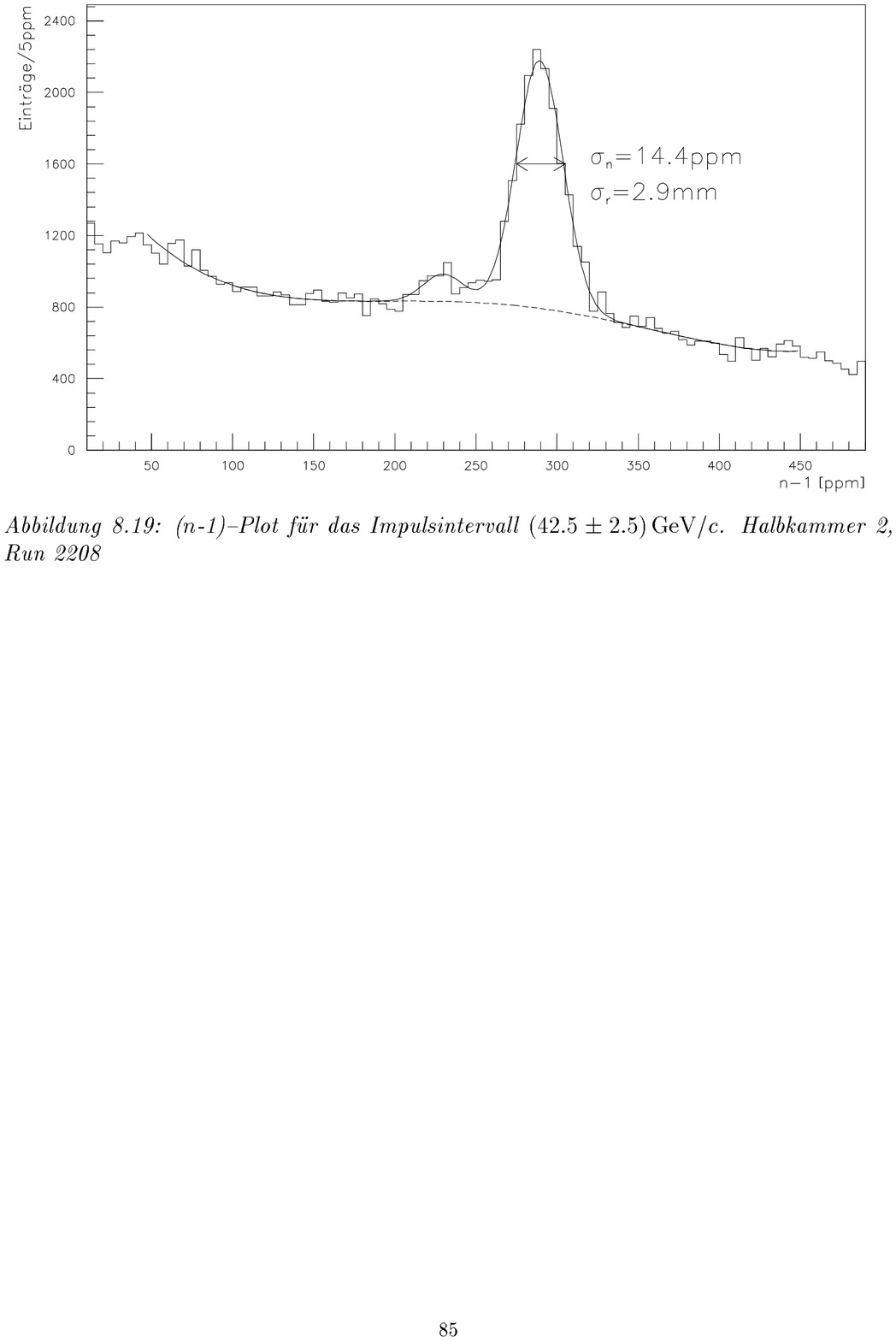}}
\caption{$n-1$ plot from WA89~\cite{Engelfried:1992aaa} for a small momentum
bin around $42.5\,\mbox{GeV}/c$. The large peak corresponds to pions and
a Gaussian fit determines the single hit resolution. The small peak at
around $230\,\mbox{ppm}$ corresponds to kaons, and the broad peak at the left
to protons close to threshold. In this plot one can also appreciate the
significant noise in the detector.}
\label{nm1plot}
\end{figure}

The sum of the single pieces has to coincide with the measured total
resolution or separation; only when the sum comes out correct the
detector system is fully understood. In the first RICH detectors the
so-called $n-1$ plot was used to calibrate and evaluate the performance
of the detectors. The procedure is the following: Predict the ring center
location using the track parameters obtained independently; assume that
all tracks a pions; assume all hits in your detector belong to this track
and calculate with equation~\ref{radformula} the apparent $n$, and fill
a histogram. An example is shown in fig.~\ref{nm1plot}.

\section{Recent Developments}

\begin{figure}[htb]
\centerline{\includegraphics[width=0.49\textwidth]{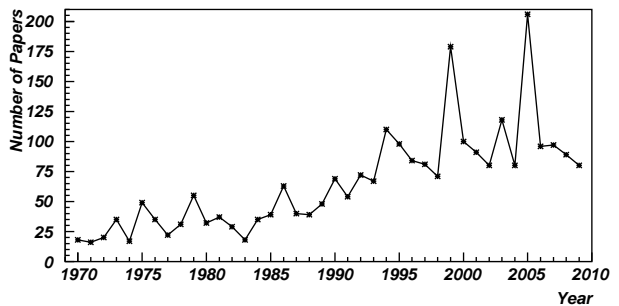}}
\caption{Number of publications (from SPIRES) about Cherenkov in general
as a function of the publication year. 2010 is only partial.}
\label{papersall}
\end{figure}

\begin{figure}[htb]
\centerline{\includegraphics[width=0.49\textwidth]{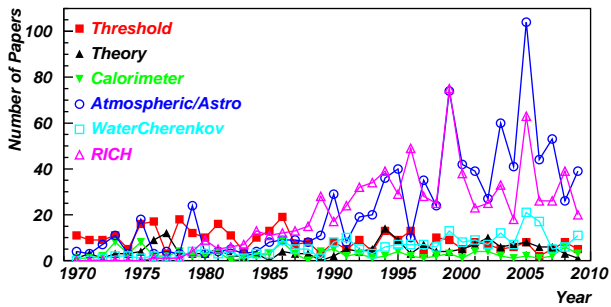}}
\caption{Same as fig.~\ref{papersall}, but separated into differemt
categories.}
\label{paperscat}
\end{figure}

Before we give a summary of new developments for Cherenkov detectors,
we will take a short detour and ask ourself if Cherenkov detectors are still
employed actively and used in new experiments.  As an indicator we
used papers contained in the 
SPIRES database\footnote{{\tt http://www.slac.stanford.edu/spires}}.
We searched for
all titles containing the words
{\sl RICH} or {\sl Ring Imaging} or {\sl cherenkov} or {\sl tscherenkov}
or {\sl cerenkov}
for every year since 1970. The results we 
divided by hand into the following Categories:\\
\indent$\bullet$ Water (and Ice) Cherenkov Detectors\\
\indent$\bullet$ Threshold (and similar like DISC) Counters\\
\indent$\bullet$ Atmospheric Cherenkov and Astronomy\\
\indent$\bullet$ Calorimeters (lead glass and similar)\\
\indent$\bullet$ Physics Results from Cherenkov detectors\\
\indent$\bullet$ Cherenkov Theory\\
\indent$\bullet$ Ring Imaging Cherenkov\\
and we excluded papers about accelerator techniques\footnote{Pavel Cherenkov
was active in this field as well.}.
As shown in fig.~\ref{papersall} the total number of publications is
increasing (with some peaky structure discussed later),
indicating that the field is still interesting and in active development. 
If we separate the publications in the different fields mentioned above
(fig.~\ref{paperscat}),
we can see that Atmospheric Cherenkov and RICH papers have the largest
numbers of publications, with the peaks in astronomy coinciding with the 
{\sl International Cosmic Ray} conferences; and the (not as prominent)
peaks for RICH papers with the RICH and other detector conferences.
The number of paper treating ``conventional'' detectors, 
like Threshold Counters, is small but constant.
Most (but not all!)\ are (highly sophisticated!)\ ``optimizations''
of the ``basics'' discussed before, but there are also new developments which
we will discuss in the following.  

The description of the new developments will be short, because all of
them have talks or even full sessions assigned at this conference.

\subsection{News on Mirrors}
The main problem to be solved for the mirrors
is still the same: they have to be thin (small $\lambda_{int}$, $X_0$) but
mechanically stable. 20~years ago CERES~\cite{Baur:1993mw}
had (one!)\ mirror made of carbon fibre, but there seems still problems
to be solved, like in the case for the LHCb Mirrors.

\subsection{News on Radiators}
Radiator gases and liquids seem to be well understood, including
Freon~\cite{Hallwell:2010}.
Aerogels are widely used, and 
some new developments are presented at this conference.
Solid radiators (for new DIRCs) are in disussion for
the detectors at the SuperBFactory and at GSI.

\subsection{News on Photon Detection}
This is the field were there are a lot of exciting developments. 
4~summary talks at this
conference~\cite{Korpar:2010,Ropelewski:2010,DallaTorre:2010,Iijima:2010}
are presenting the newest results, in addition to several contributed talks.
The newest photon detection devices are extremely fast detectors 
with high quantum efficiency, opening up new possibilites like using the
timing information for TOF identification, and suprressing noise by applying
tight timing windows to the signals.

\section{Summary}

RICHes were extensively studied and used in the last $\sim30$\,years, and are
now very well understood devices, allowing for optimizing design parameters
in a controlled way.
Their use and sophistication is still incrementing, and the field is very
active as seen from the publication study.
In particular, the newest photon detectors open new possibilities in
Cherenkov Ring Imaging.

\section*{Acknowledgment}
We would like to thank the organizers and the members of the International
Scientific Advisory Committee for the invitation to give this review.
This work was supported by CONACyT and FAI-UASLP.

\bibliographystyle{model1-num-names}
\bibliography{jurgen-rich2010.bib}

\end{document}